\documentclass[%
 reprint,
 amsmath,amssymb,
 aps,
prb,
]{revtex4-2}

\usepackage{amsmath}
\usepackage{amssymb}
\usepackage{physics}
\usepackage{graphicx}
\usepackage{dcolumn}
\usepackage{bm}
\usepackage{hyperref}
\usepackage{xcolor}

\newcommand{\bd}{\mathbf{d}}
\newcommand{\br}{\mathbf{r}}
\newcommand{\bq}{\mathbf{q}}
\newcommand{\bp}{\mathbf{p}}
\newcommand{\bk}{\mathbf{k}}

\begin{document}


\title{Bogolon-mediated electromagnetic wave absorption in multicomponent Bose-Einstein condensates}

\author{Xinyu Zhu}
\affiliation{School of Physics and Optoelectronic Engineering, Beijing University of Technology, Beijing, China, 100124}
\author{Meng Sun}
\email{msun\_89@bjut.edu.cn}
\affiliation{School of Physics and Optoelectronic Engineering, Beijing University of Technology, Beijing, China, 100124}

\date{\today}
             
\begin{abstract}
We investigate the electromagnetic wave absorption process in a coherently coupled two-component Bose-Einstein condensate model in different dimensionality at zero temperature.
As the analogue of phonon in the solid state physics, the elementary excitation of the Bose-Einstein condensate is described by Bogoliubov quasiparticle or bogolon for short.
Due to the small magnitude of the sound velocity of the bogolon, the absorption process is prohibited by the conservation of energy and momentum.
To surmount this depression, the additional degree of freedom must be considered inside of the simple Bose gas model.
In this article, we develop a microscopical theory for electromagnetic wave absorption by a two-component Bose-Einstein condensate and investigate the absorption rate dependence in different dimensions.
Our calculation shows the possibility of manipulating the absorption property by tuning the parameters of the condensates. 

\end{abstract}


\maketitle

\section{Introduction}
The experimental realization of Bose-Einstein condensate (BEC) in dilute atomic gases\cite{anderson1995,davis1995} and quasiparticles like exciton-polaritons\cite{kasprzak2006,deng2010} has triggered immense interest in the field of cold atom and light-matter coupling physics. 
From the practical point of view, the interaction between bosonic particles in condensation and cavity photons provides a valuable platform for quantum information processing\cite{brennecke2007,vanenk2004,xu2020} and quantum simulating\cite{ghosh2020,tao2022}.
On the theoretical side, the Bose-Einstein condensation itself has several fundamental questions.
In many particle physics, the elementary excitation or the quasiparticle plays a crucial role in understanding the low-energy excitation of the system. 
Recent works include the novel spectrum of elementary excitation in BEC with the Rabi and the spin-orbit coupling effect\cite{shelykh2006,shelykh2009,ravisankar2021}, the dissipation of the quasiparticles in BEC\cite{katz2002,wu2018a,wu2024}, and quasiparticle mediated interactions\cite{sun2019,villegas2019,sun2021a,sun2021b,sun2021c,cominotti2024}. 

On the other hand, radiation pressure is a phenomenon that describes the momentum transfer between light and matter\cite{gibson1970}. 
The importance of the radiation pressure-related techniques cannot be overestimated. 
In the cold atom field, it provides the theoretical basis for manipulating and trapping the particles\cite{raab1987}.
Such technique further develops the laser cooling method\cite{cohen-tannoudji1998,lett1988}, which is utilized in the formation of atomic Bose-Einstein condensate\cite{anderson1995,davis1995}.
Generally speaking, there are two types of processes for transferring energy and momentum from light to matter: light scattering and light absorption.
However, if the system is in the BEC state, the absorption process can be significantly depressed.
The reason for this decline is due to the Bogoliubov quasiparticle\cite{ozeri2005} (bogolon), the elementary excitation of Bose-Einstein condensate in weakly interacting Bose model, has a linear dispersion spectrum, and its sound velocity is much smaller than the speed of light.
Thus, the absorption process is prohibited because of the violation of the conversation laws.

\begin{figure}[b]
    \centering
    \includegraphics[width=0.75\linewidth]{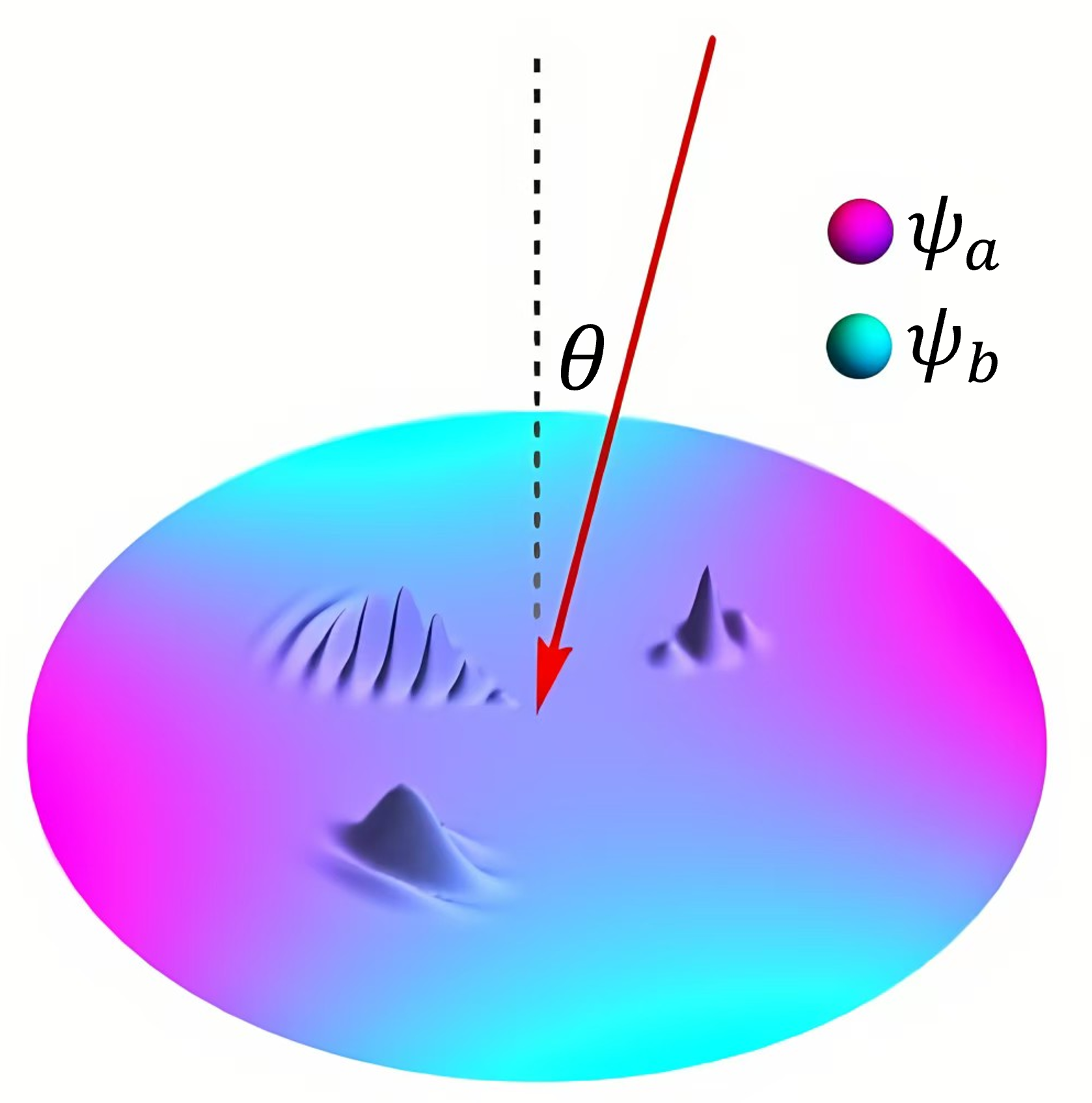}
    \caption{Schematic diagram. The electromagnetic wave (red) is approximately perpendicular ($\theta \approx 0^\circ$) shading onto the two-component Bose-Einstein condensate. The magenta and cyan colours represent $a$ and $b$ components of the Bose gas, respectively. 
    The possible excitations are indicated by the ripples on the Bose gas.}
    \label{fig:1}
\end{figure}

In this work, we consider an alternative way to enhance this absent absorption.
As shown in Fig.~\ref{fig:1}, we consider a coherently coupled two-component Bose-Einstein condensate and let the electromagnetic field nearly perpendicular shading on it.
Different from the early work\cite{ko2023}, which considers the internal degree of freedom of Bose particle, we found two new absorbing channels, and the previously opened channels are closed in this new system. 
Applying the Fermi golden rule, we further numerically calculate the absorption rate by considering the Bose gas in different dimensions.
Our work is organized as follows: In Sec. II, we introduce the Hamiltonian of the coherently coupled two-component Bose-Einstein gas model and discuss its basic properties.
In Sec. III, we discuss the Bogoliubov transformation in this model.
Sec. IV shows the possible absorbing channels and the corresponding transition rate for different configurations.
Finally, we summarize our conclusion in Sec. V.

\section{The two-component Bose-Einstein gas model}
In this work, we consider a coherently coupled two-component Bose-Einstein gas model in zero temperature. 
The Hamiltonian of the Bose system is: ($\hbar=1$)
\begin{align}
\label{eq:1-1}
\hat{H}_0&=\int d\br \left\{-\hat{\psi}_a ^\dagger (\br)\frac{\nabla^2}{2m}\hat {\psi}_a(\br)-\hat{\psi}_b ^\dagger (\br)\frac{\nabla^2}{2m}\hat {\psi}_b(\br)\right.\notag\\
&\left.+\frac{U_0}{2}\left[\hat{\psi }_a^\dagger (\br)\hat{\psi}_a^\dagger (\br)\hat{\psi}_a(\br)\hat{\psi}_a(\br)\right.\right.\notag\\
&\left.\left.+\hat{\psi }_b^\dagger (\br)\hat{\psi}_b^\dagger (\br)\hat{\psi}_b(\br)\hat{\psi}_b(\br)\right ]\right.\notag\\
&\left.+ U_1\hat{\psi }_a^\dagger (\br)\hat{\psi}_b^\dagger (\br)\hat{\psi}_a(\br)\hat{\psi}_b(\br) \right.\notag\\
&\left.+\frac{\Omega}{2} \left[ \hat{\psi}_a^\dagger \left(\mathbf{r}\right) \hat{\psi}_b\left( \mathbf{r}\right) + \hat{\psi}_b^\dagger \left(\mathbf{r}\right) \hat{\psi}_a\left( \mathbf{r}\right) \right] \right\}, 
\end{align}
where $\hat{\psi}_\eta^\dagger$ with $\eta=a,b$ are the creation operator of $a$ and $b$ component of the Bose gas (different polarization, for example), respectively; 
$U_0 $ is the intra-component interaction;
$U_1 $ is the inter-component interaction;
$\Omega = |\Omega|e^{i\phi_\Omega}$ is the coherent coupling between two condensates, and we will choose the gauge such that $\phi_\Omega=0$ in the following text.
As early theoretical and experimental works suggested\cite{wu2024,abad2013,zibold2010,recati2019,search2001}, the Hamiltonian in \eqref{eq:1-1} has $U\left(1\right) \times \mathbb{Z}_2$ symmetry, where $U\left(1\right)$ corresponds to conservation of the total number of particles, and $\mathbb{Z}_2$ corresponds to the interchange of these two components. 
Defining the total condensate density $n$ and the density difference between two components $n_d=n_a-n_b$, the ground state exhibited a paramagnetic phase $n_d=0$ when $U_1<U_0 + \frac{\abs{\Omega}}{n}$, which the $\mathbb{Z}_2$ symmetry is preserved. 
On the other hand, defining $\Delta=n\left(U_0-U_1\right)$, the ground state shows a doubly degenerate ferromagnetic phase with $n_d=\pm n \sqrt{1- \left(\frac{\abs{\Omega}}{\Delta}\right)^2} $ when $U_1>U_0+\frac{\abs{\Omega}}{n}$, which corresponds to $\mathbb{Z}_2$ breaking phase.
In this work, we will focus on the case with $\mathbb{Z}_2$ symmetry.



For further analysis, we apply the plane wave ansatz, $\hat{\psi}_\eta \left(\mathbf{r}\right) =\frac{1}{\sqrt{V}}\sum e^{i\mathbf{q} \mathbf{r}} \hat{a}_{\mathbf{q},\eta}$, and Bogoliubov approximation $\hat{a}^\dagger_{\mathbf{0},\eta}=\hat{a}_{\mathbf{0},\eta}\approx \sqrt{n_\eta}= \sqrt{\frac{n}{2}}$ where $\hat{a}_\mathbf{q,\eta}$ is the corresponding operator in momentum representation.
Then our Hamiltonian \eqref{eq:1-1} becomes:
%
\begin{align}
    \hat{H}_0&=\sum_{\bp\neq 0,\eta}^\prime\left[\varepsilon_\bp + \delta -\mu \right]\left(\hat{a}_{\bp,\eta}^\dagger \hat{a}_{\bp,\eta}+\hat{a}_{-\bp,\eta}^\dagger \hat{a}_{-\bp,\eta}\right) \notag \\
    &+\frac{n}{2} U_0 \left( \hat{a}_{\bp,\eta}^\dagger \hat{a}_{-\bp,\eta}^\dagger +\hat{a}_{\bp,\eta}\hat{a}_{-\bp,\eta} \right) \notag \\ 
    &-U_1\frac{n}{2}\left( \hat{a}_{\bp,a}^\dagger \hat{a}_{\bp,b} +\hat{a}_{\bp,a}^\dagger \hat{a}_{-\bp,b}^\dagger \right. \notag\\
    &+\left. \hat{a}_{-\bp,a}^\dagger\hat{a}_{\bp,b}^\dagger +  \hat{a}_{-\bp,a}^\dagger \hat{a}_{-\bp,b} +\mathrm{H.C.} \right) \notag \\
    &+\frac{\abs{\Omega}}{2} \left( \hat{a}_{\bp,\eta}^\dagger \hat{a}_{\bp,\bar{\eta}} + \hat{a}_{-\bp,\eta}^\dagger \hat{a}_{-\bp,\bar{\eta}} \right) , \label{eq:1-2}
\end{align}
where $\delta=\frac{n}{2}\left( 2U_0 + U_1 \right)$ and $ \mu = \frac{1}{2} \left( U_0 n + U_1 n -\abs{\Omega} \right) $ is the chemical potential and $\varepsilon_\bp = \frac{\abs{\bp}^2}{2m}$ is the kinetic energy for the Bose gas. 
The prime on the summation indicates that it is to be taken only over one-half of momentum space since the terms corresponding to $\bp$ and $-\bp$ must be counted only once. 
In Eq.~\eqref{eq:1-2}, we have neglected the homogeneous contribution from the condensed term.

\section{Bogoliubov transformation and quasi-particle}
\begin{figure*}[t]
\centering
\includegraphics[width=1\linewidth]{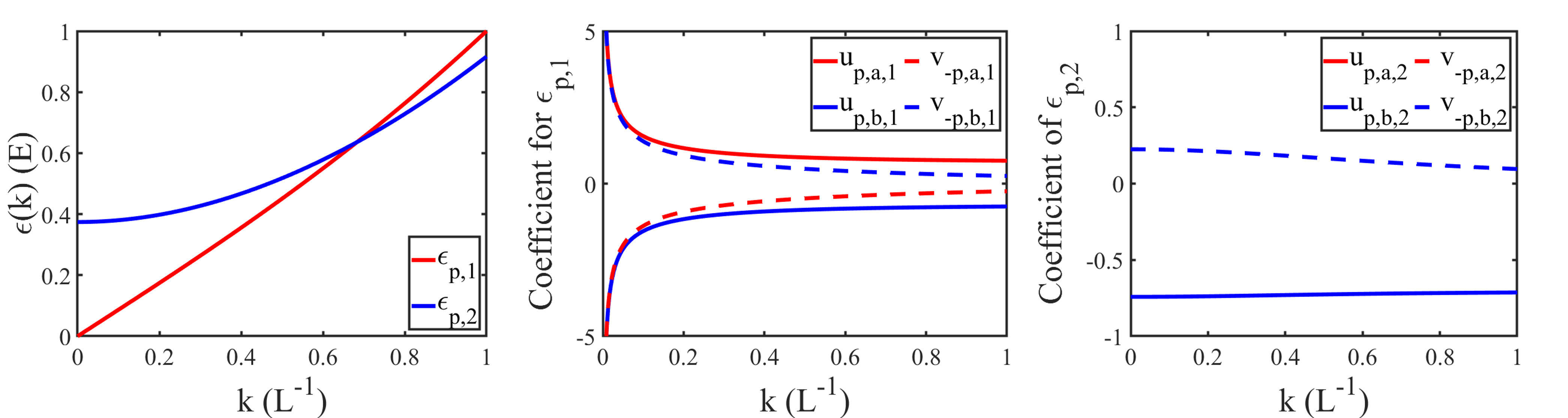}
\caption{The typical Bogoliubov spectrum and amplitude. Left: dispersion relation $\epsilon\left(k\right)$ for the case with $\mathbb{Z}_2$ symmetry. Amplitude of the Bogoliubov coefficients $\epsilon_{\bp,1}$ (middle) and $\epsilon_{\bp,2}$ (right). The parameters used are : $\abs{\Omega} = 0.2 , nU_0 = 1 , nU_1 = 0.5$.}
\label{fig:2}
\end{figure*}
The Bogoliubov transformation\cite{bogolubov2009} is a canonical transformation to diagonalize the Hamiltonian by preserving the commutation relationship. 
Initially introduced in the context of liquid helium, this technique turns out to be very fruitful and is extensively used in condensed matter physics. 

Given our Hamiltonian \eqref{eq:1-2} can be represented by the following general form
\begin{equation}
H = \sum_{\alpha \beta} A_{\alpha \beta} \hat{b}^\dagger_\alpha \hat{b}_\beta + \frac{1}{2} \sum_{\alpha \beta} B_{\alpha \beta} \hat{b}_\alpha^\dagger \hat{b}_\beta^\dagger + \frac{1}{2} \sum_{\alpha \beta} B_{\alpha \beta}^* \hat{b}_\alpha \hat{b}_\beta \label{eq:1-3}
\end{equation}
with the notation $\hat{b} = \left(\hat{a}_{\bp,a},\hat{a}_{-\bp,a},\hat{a}_{\bp,b},\hat{a}_{-\bp,b}\right)^T$. 
One can apply the Bogoliubov transformation to convert \eqref{eq:1-2} into the diagonal form, $H = \sum_{\mu}\epsilon_\mu \hat{\xi}^\dagger_\mu \hat{\xi}_\mu$, with the following transformation 
\begin{align}
\label{eq:1-5}
\hat{b}_\alpha =& \sum_\mu \left( u_{ \alpha,\mu} \hat{\xi}_{\mu} + v_{ \alpha,\mu} \hat{\xi}^\dagger_{\mu} \right),\\
\label{eq:1-6}
b_\alpha^\dagger =& \sum_\mu \left( u_{\alpha,\mu} \hat{\xi}_{ \mu}^\dagger + v_{\alpha,\mu} \hat{\xi}_{ \mu} \right),
\end{align}
where $\hat{\xi}=\left(\hat{\xi}_{\bp,1},\hat{\xi}_{-\bp,1},\hat{\xi}_{\bp,2}, \hat{\xi}_{-\bp,2} \right)^T$ are the annihilation operator for Bogoliubov quasi-particle (bogolon) from branch $1$ or $2$ with momentum $\pm\mathbf{p}$; and $u_{\alpha,\mu}$ and $v_{ \alpha,\mu}$ are the Bogoliubov coefficient. 
In our consideration, $U_1 < U_0 + \frac{\Omega}{n}$, the analytical expression of the Bogoliubov spectrum and amplitudes are\cite{tommasini2003,wu2018a,wu2024,bogolubov2009}
\begin{align}
\label{eq:1-7}
\epsilon_{\bp,1} =& \sqrt{\varepsilon_\bp\left[\varepsilon_\bp + \left(U_0+U_1\right)n\right]}, \\
\label{eq:1-8}
\epsilon_{\bp,2} =&\sqrt{\left(\varepsilon_\bp + \Omega\right)\left[ \varepsilon_\bp + \Omega + \left(U_0-U_1\right)n \right]}, 
\end{align}
and 
\begin{align}
\label{eq:1-9}
u_{\bp a,\bp 1} =& -u_{\bp b,\bp 1} = \frac{1}{\sqrt{8}} \left( \sqrt{\frac{\varepsilon_\bp}{\epsilon_{\bp,1}}}+ \sqrt{\frac{\epsilon_{\bp,1}}{\varepsilon_\bp}} \right) \\
\label{eq:1-10}
v_{\bp a,-\bp 1} =& -v_{\bp b,-\bp 1} = \frac{1}{\sqrt{8}} \left( \sqrt{\frac{\varepsilon_\bp}{\epsilon_{\bp,1}}}- \sqrt{\frac{\epsilon_{\bp,1}}{\varepsilon_\bp}} \right) \\
\label{eq:1-11}
u_{\bp a,\bp 2} =& u_{\bp b,\bp 2} = \frac{-1}{\sqrt{8}} \left( \sqrt{\frac{\varepsilon_\bp+\Omega}{\epsilon_{\bp,2}}} + \sqrt{\frac{\epsilon_{\bp,2}}{\varepsilon_\bp + \Omega}} \right) \\
\label{eq:1-12}
v_{\bp a,-\bp 2} =& v_{\bp b,-\bp 2} = \frac{-1}{\sqrt{8}} \left( \sqrt{\frac{\varepsilon_\bp+\Omega}{\epsilon_{\bp,2}}} - \sqrt{\frac{\epsilon_{\bp,2}}{\varepsilon_\bp + \Omega}} \right) 
\end{align}
where $\epsilon_{\bp,1\left(2\right)}$ are the lower (upper) branch of Bogoliubov spectrum; 
the indexes of the non-zero Bogoliubov coefficient $u_{\bp\eta,\bp\gamma}$ and $v_{\bp\eta,-\bp\gamma}$ are the momentum for Boson particle, the component of Bose gas $\eta=a,b$, the momentum for bogolon and branch of the Bogoliubov spectrum $\gamma=1,2$, respectively. 
Given the coefficients are only dependent on the magnitude of the momentum, we will neglect one of the momentum indexes for short.
As usual, we define the sound velocity $s=\sqrt{\frac{\left(U_0 +U_1\right)n}{2m}}$ and the corresponding healing length $\xi =\frac{1}{2m s}$.
These give us the following natural scale of energy, length and time as $\left[ E \right] \equiv \frac{\xi^{-2}}{2m}$, $\left[ L \right]\equiv \xi^{-1}$, and $\left[T\right] \equiv \left(\xi s\right)^{-1}$ applied in this work.
In Fig.~\ref{fig:2}, we show the typical result of the spectrum and coefficient in the natural unit.

\section{Electromagnetic wave absorption process}
\begin{figure*}[t]
\centering
\includegraphics[width=1\linewidth]{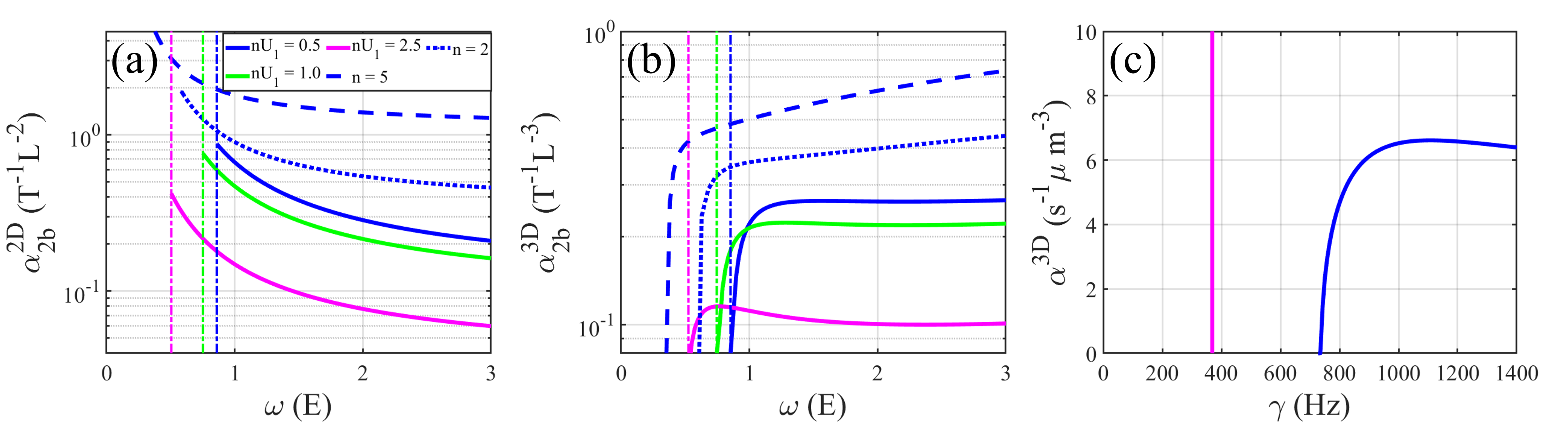}
\caption{Figure (a) and (b), the absorption rates for 2D and 3D cases. The parameters used are $\abs{\bd \cdot \hat{E}_0}=\abs{\Omega}=1$. In the solid line case, the density $n=1$ and the interaction strength $nU_0=5$ are fixed. The colour indicates different interaction strengths with $nU_1=0.5$ (blue), $nU_1=1.0$ (green), and $nU_1=2.5$ (magenta). The line style for the blue curve shows the dependency of absorption rate with different densities: $n=1$ (solid),$n=2$ (dotted) and $n=5$ (dashed).  The thin vertical dashed lines indicate the threshold frequency $\omega_c$ for the case $n=1$. Figure (c) shows the absorption result from the ultracold atom system composed of $^{23}$Na with different spin states. The magenta curve is the first-order process of $\hat{V}_{1b}$. The blue curve starts at $\gamma\approx738.36$Hz is the contribution of $\hat{V}_{2b}$. The plasma frequency is $\omega_{pl}\approx369.18$Hz.}
\label{fig:3}
\end{figure*}
Now, let us consider a weak electromagnetic field with nearly perpendicular shading on the Bose gas system. 
The whole Hamiltonian can be written as $\hat{H}=\hat{H}_0 + \hat{V}$, in which the interaction between bosons and electromagnetic wave is considered as a simple dipole interaction form\cite{kovalev2017b,kovalev2018c,ko2023}:
\begin{equation}
\label{eq:1-13}
\hat{V} = -\hat{\bd} \cdot \hat{\mathbf{E}} = - \sum_{\eta \eta'} \mathbf{d}_{\eta,\eta'} \int d \mathbf{r} \psi^\dagger_{\eta'} \left( \mathbf{r},t \right) \hat{\mathbf{E}} \psi_\eta \left( \mathbf{r},t \right). 
\end{equation}
Here, we consider the monochrome wave as a classical field as $\hat{\mathbf{E}} = \hat{\mathbf{E}}_0 e^{i \left(\mathbf{k} \mathbf{r} - \omega t \right)} + \mathrm{C. C}$. 
For simplicity, we further assume the dipole moment is the same for different components, i.e., $\mathbf{d}_{\eta'\eta}=\mathbf{d}$. 
With Fourier transformation and Bogoliubov transformation given in \eqref{eq:1-5} and $\eqref{eq:1-6}$, we can decompose the interaction term by the number of Bogoliubov quasi-particles.
For the single bogolon process, the interaction reads
\begin{align}
    \label{eq:1-14}
    \hat{V}_{1b}&=-2 \bd \cdot \hat{\mathbf{E}}_0 \sqrt{\frac{n}{2}} \notag \\
    \times&\left[ \left( v_{\bk,a,1}+u_{\bk,a,1}+v_{\bk,b,1}+u_{\bk,b,1} \right) \xi_{\bk,1}^\dagger \right. \notag\\
    &\left.+ \left(v_{\bk,a,2}+u_{\bk,a,2}+v_{\bk,b,2}+u_{\bk,b,2} \right) \xi_{\bk,2}^\dagger \right].
\end{align}
Similarly, for the two-bogolon process, we have
\begin{align}
\label{eq:1-15}
\hat{V}_{2b}&=-4\mathbf{d}\cdot \hat{\mathbf{E}}_0\sum_{\bp,\bp^\prime }\delta \left(\mathbf{p^\prime}+\mathbf{p}-\mathbf{k} \right)  \notag \\
&\times\left[ \left(u_{\bp^\prime,a,1}v_{\bp,a,1}+u_{\bp^\prime,a,1}v_{\bp,b,1} \right. \right. \notag \\
&+ \left. \left. u_{\bp^\prime,b,1}v_{\bp,a,1}+u_{\bp^\prime,b,1}v_{\bp,b,1}\right) \xi_{\bp,1}^\dagger \xi_{\bp^\prime,1}^\dagger \right.\notag\\
&\left.+ \left(u_{\bp^\prime,a,2}v_{\bp,a,2}+u_{\bp^\prime,a,2}v_{\bp,b,2} \right. \right. \notag \\
&+ \left. \left. u_{\bp^\prime,b,2}v_{\bp,a,2}+u_{\bp^\prime,b,2}v_{\bp,b,2} \right) \xi_{\bp,2}^\dagger \xi_{\bp^\prime,2}^\dagger \right. \notag\\
&+\left. \left( u_{\bp^\prime,a,1}v_{\bp,a,2}+u_{\bp^\prime,a,1}v_{\bp,b,2} \right. \right. \notag \\
&+ \left. \left. u_{\bp^\prime,b,1}v_{\bp,a,2}+u_{\bp^\prime,b,1}v_{\bp,b,2} \right. \right. \notag \\
&+ \left. \left. u_{\bp,a,2}v_{\bp^\prime,a,1}+u_{\bp,a,2}v_{\bp^\prime,b,1} \right. \right. \notag \\
&+ \left. \left. u_{\bp,b,2}v_{\bp^\prime,a,1}+u_{\bp,b,2}v_{\bp^\prime,b,1} \right)\xi_{\bp^\prime,1}^\dagger \xi_{\bp,2}^\dagger\right]. 
\end{align}
Here are some comments about the interaction terms\eqref{eq:1-14} and \eqref{eq:1-15}.
\textbf{(i).} In this work, we consider the Bose gas in the zero temperature limit.
Then, the process accompanied by the emission of bogolons is considered exclusively.
\textbf{(ii).} For the same reason, we only count the electromagnetic wave absorption processes and disregard the term containing $\sim \hat{\mathbf{E}}^\dagger_0$.
\textbf{(iii).}  The $\hat{V}_{1b}$ term describes the emission of single bogolon to branch $\gamma=1$ or $\gamma=2$, and the $\hat{V}_{2b}$ term describes the emission of double bogolon spontaneously to different branches.

We apply the Fermi golden rule to calculate the absorption rate.
The absorption probability for different interaction channels are
\begin{equation}
\label{eq:1-16}
\alpha_{i\rightarrow f} = \frac{2\pi}{\hbar} \abs{\bra{f} \hat{V} \ket{i}}^2 \delta\left( E_f - E_i -\omega \right), 
\end{equation}
where $\ket{i}$ is the initial state (all particles are in the BEC state in the zero temperature limit).
The final states $\ket{f}$ depend on the form of the interaction in Eq.~\eqref{eq:1-14} and \eqref{eq:1-15}.
For the single bogolon process, the final state is a single Bogoliubov quasiparticle emitted from the BEC by interaction $\hat{V}_{1b}$, and the transition rate is 
\begin{align}
\label{eq:1-17}
\alpha_{1b} = \frac{64\pi n}{\hbar} \abs{\bd \cdot \hat{\mathbf{E}}_0}^2 \abs{u_{\bk,a,2}+v_{\bk,a,2}}^2 \delta\left( \epsilon_{\bk,2} - \omega_\bk \right).  \end{align}
For the single bogolon process, given that the sound velocity of bogolon is much smaller than the speed of light $s \ll c$, we can neglect the contribution which the bogolon scattered into the lower branch of the spectrum due to the conservation of momentum and energy.
However, the absorption is permitted by emitting a single bogolon from the upper branch because of the finite gap of this excitation, as shown in Eq.~\eqref{eq:1-8} and Fig.~\ref{fig:2}.
According to the size of the band gap~\eqref{eq:1-20}, we can estimated the peak of the absorption, $\omega_c^{1b},$ happens slightly above the plasma frequency $\omega_{pl}\equiv\sqrt{\Omega\left(\Omega+\Delta\right)}$~\cite{cominotti2024}.

For the double-bogolon process, the transition rate reads
\begin{equation}
\label{eq:1-18}
\alpha_{2b} = \alpha_{2b}^{11} + \alpha_{2b}^{22} + \alpha_{2b}^{12}, 
\end{equation}
where $\alpha_{2b}^{11}$ ($\alpha_{2b}^{22}$) describes the contribution in which the two bogolons are both scattered into the lower (upper) Bogoliubov branch and $\alpha_{2b}^{12}$ describes the contribution which the two bogolons are scattered into the lower and upper branch each 
\begin{align}
\label{eq:a11}
    \alpha_{2b}^{11} =& \frac{32\pi}{\hbar} \abs{\mathbf{d} \cdot \hat{E}_0}^2 \sum_\bp \delta\left(\epsilon_{\bp+\bk,1}+\epsilon_{-\bp,1}-\omega_\bk\right) \notag \\
    \times & \abs{\left(u_{\bp+\bk,a,1}+u_{\bp+\bk,b,1}\right)\left(v_{\bp,a,1}+v_{\bp,b,1}\right)}^2,\\
\label{eq:a22}
    \alpha_{2b}^{22} =& \frac{32\pi}{\hbar} \abs{\mathbf{d} \cdot \hat{E}_0}^2 \sum_\bp \delta\left(\epsilon_{\bp+\bk,2}+\epsilon_{-\bp,2}-\omega_\bk\right) \notag \\
    \times & \abs{\left(u_{\bp+\bk,a,2}+u_{\bp+\bk,b,2}\right) \left( v_{\bp,a,2}+v_{\bp,b,2}\right)}^2, \\
\label{eq:a12}
    \alpha_{2b}^{12} =& \frac{32\pi}{\hbar} \abs{\mathbf{d} \cdot \hat{E}_0}^2 \sum_\bp \delta\left(\epsilon_{\bp+\bk,1}+\epsilon_{-\bp,2}-\omega_\bk\right) \notag\\
    \times& |\left(u_{\bp+\bk,a,1}+u_{\bp+\bk,b,1}\right)\left(v_{\bp,a,2}+v_{\bp,b,2}\right)\\ +&\left(u_{\bp,a,2}+u_{\bp,b,2}\right)\left(v_{\bp+\bk,a,1}+v_{\bp+\bk,b,1}\right)|^2.\notag
\end{align}
Noticing the antisymmetry of Bogoliubov coefficients of lower branch in \eqref{eq:1-9} - \eqref{eq:1-10}, we have $u_{\bp,a,1}+u_{\bp,b,1}=0$.
Then we can conclude that $\alpha_{2b}^{11} = \alpha_{2b}^{12} = 0$ and the only non-zero contribution is $\alpha_{2b}^{22}$
\begin{align}
\label{eq:1-19}
\alpha_{2b}=&\frac{512}{\hbar}\abs{\bd \cdot \hat{E}_0}^2 \int \frac{d\bp}{\left(2\pi\right)^D} \delta(\epsilon_{\bp+\bk,2}+\epsilon_{-\bp,2}-\omega_\bk) \notag \\
\times & \abs{u_{\bp+\bk,a,2} v_{\bp,a,2}}^2, 
\end{align}
where $D$ is the dimensionality of the Bose gas.

Before the discussion about the numerical result, let us compare the possible absorption channels with previous work\cite{ko2023}.
When the internal degree of freedom for the Bose particle is considered, unlike the two-component Bose gas model, the Bogoliubov spectrum and coefficient are unique.
At small $\bp \ll \xi^{-1}$, its spectrum is linear and gapless, which is similar to $\epsilon_{\bp,1}$ in Eq.~\eqref{eq:1-7}.
Thus, we can conclude that the coherently coupled two-component model opens a new absorption channel because of the finite gap of $\epsilon_{\bp,2}$ in Eq.~\eqref{eq:1-8}.
Moreover, the model of Bose gas with the internal degree of freedom also provides the double-bogolon absorption channel.
Similar to $\alpha_{2b}^{11}$ in Eq.~\eqref{eq:a11}, the electromagnetic wave is absorbed by emitting two bogolons to the linear spectrum.
However, due to the symmetry property of the Bogoliubov coefficient, the absorption channel to the $\epsilon_{\bp,1}$ branch is closed in the two-component Bose gas model.
At last, we want to point out that the new absorption channel in Eq.~\eqref{eq:a22} bears a different excitation spectrum and Bogoliubov coefficient, which results in a novel absorption dependence.

By assuming the small incident angle of electromagnetic wave, we approximate the Bogoliubov spectrum \eqref{eq:1-8} up to the second order for further calculation.
In Figs.~\ref{fig:3}(a) and (b), we show the numerical result of absorption rate for different dimensionalities as the function of electromagnetic wave frequency for the double-bogolon process (see Appendix \ref{ap:2} for calculation details).
The colour indicates the absorption result for different interaction strength $nU_1$ by fixing the parameters $\abs{\Omega}=1$ and $nU_0=5$.
The line style represents the result with different condensed densities: $n=1$ (solid line), $n=2$ (dotted line) and $n=5$ (dashed line).
The vertical dashed lines in each figure indicate the threshold frequency of the absorption for two bogolon process (only the case with $n=1$ is plotted for clearness).
A simple estimation in \eqref{eq:A-3} and \eqref{eq:A-2} for this threshold frequency leads to $\omega_c^{2b}\approx 2\omega_c^{1b} > 2\omega_{pl}$.
In general, by fixing the wave frequency above the threshold $\omega_c^{2b}$, one can increase the absorption rate by decreasing the inter-component interaction strength $U_1$ or by increasing the condensed density $n$ as shown in Fig.~\ref{fig:3} (a) and (b).

The dimensionality of the Bose gas also affects the absorption behaviour significantly.
For the two-dimensional case, we find the absorption rates decline as a function of the electromagnetic wave frequency above the threshold.
Furthermore, as suggested in Eq.~\eqref{eq:b-11}, the absorption rate is finite at the threshold. 
For the three-dimensional Bose gas model, we find that the absorption rate behaves similar to a Heaviside step function.

In order to observe the absorption phenomenon experimentally,
one can consider the spin mixtures of ultra-cold atom systems.
For specific case, we can consider the sodium ($^{23}$Na) atoms\cite{cominotti2022} with a mass of $m\approx22.9\,\text{u}$, $\Delta/h = 450\,\text{Hz}$, and $\Omega/\left(2\pi h\right)= 33\,\text{Hz}$.
Considering the dipole interaction $\abs{\mathbf{d}\cdot \hat{E}_0}/h=0.1\,\omega_{pl}$ and the density of the sodium $n=10\,\mu \text{m}^{-3}$, we calculate the real absorption rate in Fig.~\ref{fig:3}(c).
For the single bogolon process (magenta curve), the absorption behaves like a delta function which located slightly above the plasma frequency $\omega_{pl}$.
On the other hand, the contribution from the two-bogolon process starts slightly above $2\omega_{pl}$ as shown in Fig.~\ref{fig:3} (c) by the blue curve.
From the analysis above, one can conclude that the contributions of the two term $\hat{V}_{1b}$ and $\hat{V}_{2b}$ are well separated in the frequency domain.

Furthermore, it should be noted that the two-bogolon process originating from $\hat{V}_{2b}$ is dominant within its corresponding frequency domain.
Given the interaction $\hat{V}_{1b}$ is proportional to the condensed density, one can anticipate its interaction strength is the order of magnitude larger than that of $\hat{V}_{2b}$.
This naturally raises the question of whether the second-order perturbation of $\hat{V}_{1b}$ would domain the two-bogolon process.
To clarify this question, we consider the Fermi golden rule to the second-order~\cite{GROSSO2000425},
\begin{align}
    \label{eq:add-1}
    \alpha^{\left(2\right)}_{i\rightarrow f}=&\frac{2\pi}{\hbar}\abs{\sum_m \frac{\bra{f}\hat{V}\ket{m} \bra{m}\hat{V}\ket{i}}{E_i+\omega-E_m}}^2 
     \delta\left(E_f-E_i-2\omega \right)
\end{align}
where $\ket{m}$ is the intermediate state which is unfettered by conservation law.
For this specific question only, we have $\hat{V}=\hat{V}_{1b}$;
The initial state is the state with all particles in the BEC state as usual and the final state $\ket{f}$ which must satisfied the conservation law, we can consider the state with two emitting bogolons in upper branch $\ket{\epsilon_{\bk+\bp,2},\epsilon_{\bk-\bp,2}}$, for instance.
Due to the property of $\hat{V}_{1b}$ can create one bogolon each time, the intermediate state have to be $\ket{\epsilon_{\bk\pm \bp,2}}$.
Then, we have the following result for the two-bogolon process in the second-order of $\hat{V}_{1b}$,
\begin{align}
    \label{eq:add-2}
    \alpha_{1b}^{(2)}=&\frac{64\pi n^2}{\hbar}\abs{\bd \cdot \hat{E}_0}^4\int \frac{d\bp}{(2\pi)^D}\delta \left(\epsilon_{\bk+\bp,2}+\epsilon_{\bk-\bp,2}-2\omega_\bk \right)\nonumber\\
    &\times \abs{\left(u_{\bk+\bp,2}+v_{\bk+\bp,2}\right)\left(u_{\bk-\bp,2}+v_{\bk-\bp,2}\right)}^2\nonumber\\
    &\times\abs{\frac{1}{\omega_\bk -\epsilon_{\bk+\bp,2}}+\frac{1}{\omega_\bk-\epsilon_{\bk-\bp,2}}}^2.
\end{align}
Given the conservation of energy from $\delta-$function, we have $\alpha^{\left(2\right)}_{1b}=0$.
A similar derivation will lead to the same result, if one consider the emitting bogolons are form lower branch or lower and upper branch both.
Thus we conclude that because of the destructive interference between the intermediate states, the transition of two-bogolon process due to the second-order perturbation of $\hat{V}_{1b}$ is zero and the two-bogolon process described by \eqref{eq:1-15} is particularly significant in its frequency domain under perturbation condition.
Interestingly, a similar effect occurs in clean single-band s-wave superconductors: the optical transitions across the superconducting gap induced by uniform light are forbidden\cite{mahan2000a,parafilo2023a, parafilo2024}.
This happens because, according to the BCS theory, the electron-like and hole-like states are orthogonal to one another, resulting in matrix elements for optical absorption vanishes.
One should also notice that there exist another possible two-bogolon process in second order due to the Beliaev damping and $\hat{V}_{1b}$ which a virtual bogolon excited by $\hat{V}_{1b}$ decays into two bogolons in upper and lower branch each\cite{wu2024}.
Thus a further analysis is required, but the discussion of this process is beyond the scope of this work.
At last, given the generality of our theoretical framework, we anticipate a similar behaviour across diverse platforms, such as exciton-polariton where the plasma frequency $\omega_{pl}$ exceeds that of ultracold atomic systems by orders of magnitudes and the Bose-Einstein condensate can be treated as a quasi-two-dimensional system approximately.

\section{Conclusion}
In summary, we study the electromagnetic wave absorption process for a coherently coupled two-component Bose-Einstein condensate model in the paramagnetic phase.
Due to the symmetry (and antisymmetry) properties of the Bogoliubov coefficients, the electromagnetic wave absorption process can only happen from the upper branch of the excitation spectrum.
For the single-bogolon process by $\hat{V}_{1b}$, the absorption rate shows a simple delta function behaviour.
For the double-bogolon process by $\hat{V}_{2b}$, we calculate the absorption rate in different dimensionalities.
Although the threshold behaviour is found in all cases, the detailed properties are different case by case. 
For the two-dimensional case, the threshold shows a finite peak and decreases gradually.
In the three-dimensional case, the absorption behaves similarly to the Heaviside step function.
Moreover, we analyse the double-bogolon processes arising from the second-order perturbation of $\hat{V}_{1b}$.
Due to the symmetry of the wavefunction, we find this second-order perturbation vanishes.
Thus, our calculations indicate the contributions from $\hat{V}_{2b}$ is significant under perturbation condition in its frequency domain, which suggests a new way to observe this novel interaction.
Our finding reveals a new opportunity to manipulate the electromagnetic wave and the Bose gas in condensate, which, in principle, can be considered as a quantum gate for the incident wave by tuning the absorption threshold.

\begin{acknowledgments}
M. S. and X.Y. Zhu thanks for the insightful discussion with Dr.Anton Parafilo, Dr.Vadim Kovalev, Dr YuSong Cao and Dr. WenYu Wang. 
This work is supported by the R\&D Program of Beijing Municipal Education Commission (KM202410005011). 
X.Y. Zhu is also partially supported by the National Science Foundation of China (Grant No. NSFC-11874002).
\end{acknowledgments}

\appendix

\section{Absorption rate calculation} \label{ap:2}
For further analysis, we approximate Eq.~\eqref{eq:1-8} up to second order,
\begin{equation}
\label{eq:1-20}
\epsilon_{\bp,2} = \sqrt{\left(\varepsilon_\bp + \Omega\right) \left( \varepsilon_\bp + \Omega + \Delta \right)}\approx \zeta_0+\zeta_2 p^2
\end{equation}
where we define $\Delta = n \left( U_0 - U_1 \right)$, $\zeta_0=\sqrt{\Omega \left(\Omega+\Delta \right)}$, and $ \zeta_2=\frac{1}{4m\zeta_0 }\left(2\Omega +\Delta \right)$ for convenient.

\subsection{Two bogolon in 2D}
Now, let us consider the absorption rate Eq.~\eqref{eq:1-19} in different dimensionalities.
In the 2D case, we consider the electromagnetic wave to be nearly perpendicularly shading on the Bose gas. 
Thus, by denoting the wavevectors which are parallel and vertical to the condensate as $\bk_\parallel$ and $k_\perp$, we have $\omega_k=c\sqrt{\bk_\parallel^2+k_\perp^2}\approx=ck_\perp\equiv \omega$.
Then the absorption probability reads, 
\begin{align}
\alpha^{2D}&=\frac{128}{\hbar \pi^2}\abs{\bd \cdot \hat{E}_0}^2\int d\bp \abs{u(\bp+\bk_\parallel )v(\bp)}^2\notag\\
&\times \delta \left(\zeta_0+\zeta_2\left(\bp+\bk_\parallel \right)^2+\left(\zeta_0+\zeta_2 \bp^2\right)-\omega_\bk\right)\notag\\
&=\frac{128}{\hbar \pi^2}\abs{\bd \cdot \hat{E}_0}^2I^{2D}\label{eq:1-28}
\end{align}
Define $q=\abs{\bp +\bk_\parallel }$ and $x=q^2$, this integration becomes
\begin{align}
I^{2D}&=\int _0^\infty dp \int _{\abs{p-k_\parallel }}^{p+k_\parallel }dq \delta \left(2\zeta_0+\zeta_2p^2+\zeta_2q^2-\omega\right)\notag\\
&\times \frac{p u^2\left(\sqrt{x}\right) v^2\left(p \right)}{\sqrt{\left[ \left( p+k_\parallel \right)^2 - x \right] \left[ x - \left( p-k_\parallel \right)^2 \right]}} \notag \\
&=\frac{1}{2}\int _0^\infty dp \frac{p u^2(\sqrt{x_0})v^2(p)}{\sqrt{\left[\left(p+k_\parallel \right)^2-x_0\right]\left[x_0-\left(p-k_\parallel \right)^2\right]}}\notag\\
&\times \Theta\left(x_0-\left(p-k_\parallel \right)^2\right)\Theta \left(\left(p+k_\parallel \right)^2-x_0\right)\label{eq:1-29}
\end{align}
where 
$x_0 = \frac{\omega-\zeta_2p^2-2\zeta_0}{\zeta_2} $.
Considering the Heaviside function, we have
\begin{align}
\frac{k_{\|}-\sqrt{2 \lambda^{2}-k_{\|}^{2}}}{2}<p<\frac{k_{\|}+\sqrt{2 \lambda^{2}-k_{\|}^{2}}}{2} \label{eq:1-30} 
\end{align}
for the region $ 2 \zeta _{0}+\frac{1}{2} \zeta _{2} k_{\|}^{2}<\omega<2 \zeta _{0}+\zeta _{2} k_{\|}^{2}$ and 
\begin{align}
\frac{-k_{\|}+\sqrt{2 \lambda^{2}-k_{\|}^{2}}}{2}<p<\frac{k_{\|}+\sqrt{2 \lambda^{2}-k_{\|}^{2}}}{2} \label{eq:1-31}
\end{align}
for the region $\omega \geq 2 \zeta _{0}+\zeta _{2} k_{\|}^{2}$
where $\lambda =\sqrt{\frac{\omega-2\zeta_0}{\zeta_2}}$.
Hence, from the discussion above, we can conclude the threshold frequency in 2D is 
\begin{equation}
    \label{eq:A-3}
    \omega_c^{2b,2D} =2\zeta_0+ \frac{1}{2}\zeta_2 k_\parallel^2,
\end{equation}
which the threshold frequency is two times of the gap plus a modification by incline angle.

Applying \eqref{eq:1-30} and \eqref{eq:1-31} to the integration \eqref{eq:1-29}, we can calculate the absorption coefficient
\begin{align}
\label{eq:1-32}
\alpha_{2b}^{2D} =& \frac{64}{\hbar \pi^2}\abs{\bd \cdot \hat{E}_0}^2 \notag\\
&\times \int_{p_1}^{p_2} dp \frac{pu^2(\sqrt{x_0})v^2(p)}{\sqrt{\left[\left(p+k_\parallel \right)^2-x_0\right]\left[x_0-\left(p-k_\parallel \right)^2\right]}} 
\end{align}

Next, we would like to discuss the approximated behaviour of \eqref{eq:1-32} in the extreme case where $2\lambda^2 -k^2_\parallel \approx 0$.
That is, we want to see the behaviour of the absorption rate near the threshold frequency.
As shown in Fig.~\ref{fig:2}, the Bogoliubov coefficient is not sensitive to the changing of momentum.
Thus, we treat the Bogoliubov coefficients as constant, $u_0 \equiv u\left(\bk=0\right)$ and $v_0 \equiv v \left(\bk=0\right)$, in the following discussion and get 
\begin{align*}
I^{2D}(\omega)&\approx \frac{u_0^2 v_0^2}{2}\int_{p_1}^{p_2} dp \frac{p}{\sqrt{\left[\left(p+k_\parallel \right)^2-x_0\right]\left[x_0-\left(p-k_\parallel \right)^2\right]}}\\
&=\frac{u_0^2 v_0^2}{4}\int _{p_1}^{p_2}dp \frac{p}{\sqrt{\left(p-c\right)\left(p-d\right)\left(p-a\right)\left(b-p\right)}}
\end{align*}
with the definition
\begin{align*}
a&=p_2=\frac{1}{2}\left(k+\sqrt{\frac{2\omega -4\zeta_0}{\zeta_2}-k^2}\right)=\frac{1}{2}\left(k+\sqrt{2\lambda^2-k^2}\right)\\
b&=p_1=\frac{1}{2}\left(k-\sqrt{2\lambda^2-k^2}\right)\\
c&=\frac{1}{2}\left(-k+\sqrt{2\lambda^2-k^2}\right)\\
d&=\frac{1}{2}\left(-k-\sqrt{2\lambda^2-k^2}\right)
\end{align*}
such an integral has the analytical solution in the standard integral book \cite{gradshteyn2007}
\begin{align*}
&\int_u^a dx \frac{x}{\sqrt{\left(a-x\right)\left(x-b\right)\left(x-c\right)\left(x-d\right)}}\\
&=\frac{2}{\sqrt{\left(a-c\right)\left(b-d\right)}}\left\{\left(a-d\right)\Pi\left(\mu,\frac{b-a}{b-d},r\right)+dF\left(\mu,r\right)\right\}
\end{align*}
with the condition
\begin{align*}
a>u\ge b>c>d
\end{align*}
and the definition:
\begin{align*}
\mu &= \arcsin\sqrt{\frac{\left(b-d\right)\left(a-u\right)}{\left(a-b\right)\left(u-d\right)}}\\
r&=\sqrt{\frac{\left(a-b\right)\left(c-d\right)}{\left(a-c\right)\left(b-d\right)}}
\end{align*}
and $F\left(\mu, r\right)$ is the incomplete elliptic integral of the first kind and $\Pi\left( \mu, \nu,r\right)$ is the incomplete elliptic integral of the third kind.

\begin{figure}[t]
\centering
\includegraphics[width=0.75\linewidth]{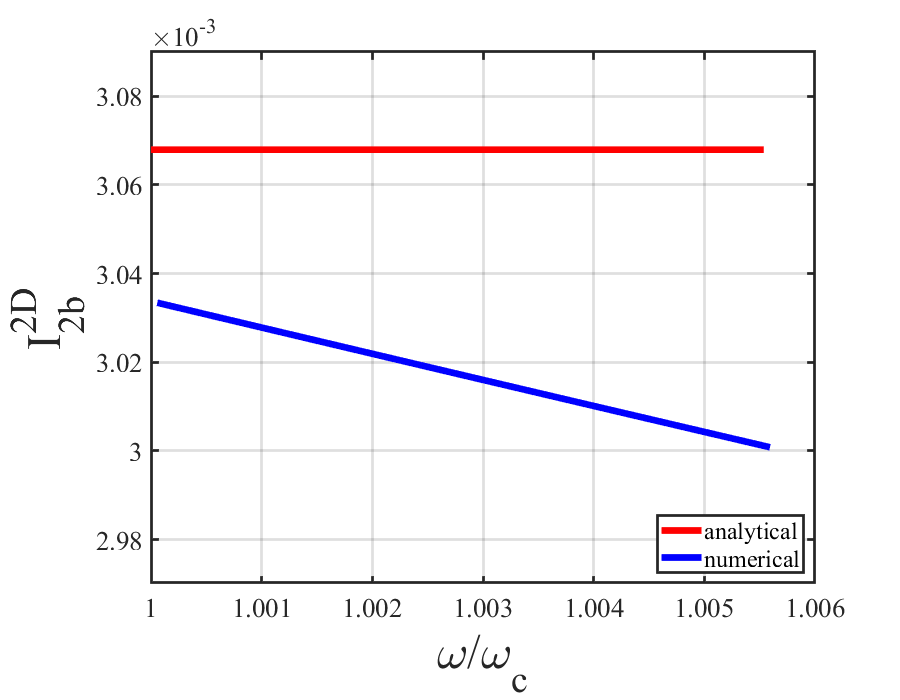}
\caption{The numerical and analytical solutions of $I^{2D}$ near the threshold frequency. The parameters are $nU_0=2nU_1=2\Omega=1$.}
\label{fig:4}
\end{figure}

In our consideration, $u=b$, we have $\mu =\arcsin 1=\frac{\pi}{2}$, so the elliptic integrals become complete.
Let $a=\frac{1}{2}\left(k+\varepsilon\right)$, we have
\begin{equation}
\label{eq:b-11}
I^{2D}\approx \frac{u_0^2 v_0^2}{2}\frac{k+\varepsilon}{k} \left\{\Pi\left(-\frac{\varepsilon}{k},\frac{\varepsilon}{k}\right)-\frac{1}{2}K\left(\frac{\varepsilon}{k}\right)\right\}
\end{equation}
We can further investigate the critical case in the limit $\varepsilon \to 0^+$.
Our analytical approximation suggests the result is  finite (notice that $k>0$)
\begin{equation}
    \label{eq:b-12}
    \lim_{\varepsilon \to 0^+} I^{2D} = \frac{\pi u^2_0 v^2_0}{4}.
\end{equation}
In Fig.~\ref{fig:4}, we show the comparison of $I^{2D}$ from numerical integration \eqref{eq:1-29} and analytical approximation \eqref{eq:b-11} near the threshold frequency.
Therefore, based on the above results, we can conclude that \( I^{2D} \) exhibits a finite solution at the threshold frequency, differing from the \( I^{1D} \) case as shown in Fig.\ref{fig:3}.

\subsection{Two bogolon in 3D}
In 3D case, we have
\begin{align}
\alpha^{3D}&=\frac{64}{\hbar \pi^3}\abs{\bd \cdot \hat{E}_0}^2\int \abs{u(\bp+\bk)v(\bp)}^2\notag\\
&\times \delta \left(\zeta_0+\zeta_2\left(\bp+\bk \right)^2+\left(\zeta_0+\zeta_2 \bp^2\right)-\omega_\bk\right)\notag\\
&=\frac{64}{\hbar \pi^2}\abs{\bd \cdot \hat{E}_0}^2I^{3D}\label{eq:1-33}
\end{align}
In the spherical coordinate system, without losing any generality, we choose the wavevector $\bk$ as the polar axis:
\begin{align}
\label{eq:I3D}
I^{3D}&=\int_0^\infty p^2dp \int _0^\pi \sin \theta d\theta \int_0^{2\pi} d\phi \abs{u(\bp+\bk)v(\bp)}^2\notag\\
&\times \delta \left(\zeta_0+\zeta_2\left(\bp+\bk \right)^2+\left(\zeta_0+\zeta_2 \bp^2\right)-\omega_\bk\right)
\end{align}
Apply the same trick by denoting $q=\abs{\bk +\bp}$, and the identity $\sin \theta =\frac{\sqrt{\left[\left(p+k\right)^2-q^2\right]\left[q^2-\left(p-k\right)^2\right]}}{2kp}$.

Then we have 
\begin{align*}
I^{3D}&=2\pi \int_0^\infty dp \int_{\abs{p-k}}^{p+k} dq\frac{qp}{k}\abs{u(\bq)v(\bp)}^2\notag\\
&\times \delta \left(\zeta_0+\zeta_2\left(\bp+\bk \right)^2+\left(\zeta_0+\zeta_2 \bp^2\right)-\omega_\bk\right)\notag\\
&=\pi \int_0^\infty dp \frac{p\abs{u(\sqrt{x_0})v(p)}^2}{\zeta_2k}
\end{align*}
where
$x_0=\frac{ck-\zeta_2p^2-2\zeta_0}{\zeta_2}$,
and with the condition
\begin{align}
(p-k)^2\le x_0\le (p+k)^2 \label{eq:1-34}
\end{align}
According to \eqref{eq:1-34} and the definition of $x_0$, we have the region of the integration
\begin{align*}
0<p_1&\le p\le p_2\notag\\
p_1&=\frac{-k+\sqrt{k^2+4\lambda}}{2}\notag\\
p_2&=\frac{k+\sqrt{k^2+4\lambda}}{2}\notag
\end{align*}
where $\lambda=\frac{ck-2\zeta_0-\zeta_2k^2}{2\zeta_2}$.
To keep the solution of $p_1$ and $p_2$ to be real, we have the condition $k^2+4\lambda >0$, which gives the limitation of incident electromagnetic wave wavevector
\begin{align*}
\frac{c-\sqrt{c^2-4\zeta_2 \zeta_0}}{\zeta_2}<k<\frac{c+\sqrt{c^2-4\zeta_2 \zeta_0}}{\zeta_2}
\end{align*}

Thus the integration of \eqref{eq:1-33} is
\begin{align}
\alpha_{2b}^{3D} = \frac{64}{\hbar \pi^2}\abs{\bd \cdot \hat{E}_0}^2 \int_{p_1}^{p_2}dp \frac{p\abs{u(\sqrt{x_0})v(p)}^2}{\zeta _2k}\Theta (k-k_0)\label{eq:1-35}
\end{align}
where
\begin{align}
k_0 = \frac{c-c\sqrt{1-\frac{8\zeta_2\zeta_0}{c^2}}}{2\zeta_2}. \label{eq:A-1}
\end{align}
The threshold frequency in 3D can be estimated by ~\eqref{eq:A-1}.
Up to second order, we have
\begin{align}
\label{eq:A-2}
\omega_c^{2b,3D}=ck_0
\approx 2\zeta_0 + \frac{4\zeta_2 \zeta_0^2}{c^2}.
\end{align} 
Thus, the threshold frequency in both 2D and 3D cases is at least twice the gap energy, a result consistent with physical intuition.


At last, let us discuss the limitation due to the parabolic approximation of the spectrum~\eqref{eq:1-20}.
Such an approximation is validate when $p< \sqrt{\left(\Delta+2\Omega
\right)2m}$.
According to \eqref{eq:1-35}, the upper limit of the integration must satisfy $p_2=\frac{k+\sqrt{k^2+4\lambda}}{2}<\sqrt{\left(\Delta+2\Omega\right)2m}$.
Then, considering the speed of electromagnetic wave is much larger than other quantities, we conclude that we can neglect higher order contribution of \eqref{eq:1-20} when the electromagnetic wave frequency $\omega<2\sqrt{\Omega\left(\Omega+\Delta\right)}+\frac{\left(2\Omega+\Delta\right)^2}{\sqrt{\Omega\left(\Omega+\Delta\right)}}$.

\bibliography{ref}

\end{document}